\documentclass[aps,twocolumn,groupedaddress,amsmath,amssymb]{revtex4}
\usepackage[pdftex]{graphicx}
\usepackage{amssymb}
\usepackage{bm}
\begin{document}
\setcounter{page}{1}
\title[]{Phantom Field from Conformal Invariance}
\author{Mokhtar Hassa\"{\i}ne}\email{hassaine-at-inst-mat.utalca.cl}
\affiliation{Instituto de Matem\'atica y F\'{\i}sica, Universidad de
Talca, Casilla 747, Talca, Chile,}
\affiliation{Centro~de~Estudios~Cient\'{\i}ficos~(CECS),
~Casilla~1469,~Valdivia,~Chile.}

\begin{abstract}
We establish a correspondence between a conformally invariant
complex scalar field action (with a conformal self-interaction
potential) and the action of a phantom scalar field minimally
coupled to gravity (with a cosmological constant). In this
correspondence, the module of the complex scalar field is used to
relate conformally the metrics of both systems while its phase is
identified with the phantom scalar field. At the level of the
equations, the correspondence allows to map solution of the
conformally non-linear Klein-Gordon equation with vanishing
energy-momentum tensor to solution of a phantom scalar field
minimally coupled to gravity with cosmological constant satisfying a
massless Klein-Gordon equation. The converse is also valid with the
advantage that it offers more possibilities owing to the freedom of
rewriting a metric as the conformal transformation of another
metric. Finally, we provide some examples of this correspondence.

\end{abstract}

\maketitle
\section{Introduction}
Conformal techniques and conformal invariance have been widely used
in general relativity, quantum field theory in curved spaces and in
string theories (for a review, see e.g. \cite{Faraoni:1998qx}). It
is well-known that the massless Klein-Gordon action for a scalar
field is conformally invariant only in two dimensions while in
higher dimensions, the conformal invariance can be achieved by
introducing a nonminimal coupling \cite{Callan:1970ze}. In this
paper, we precisely consider the conformally invariant Klein-Gordon
action for a complex scalar field $\Psi$ in arbitrary dimension with
a potential term that does not spoil the conformal invariance. We
restrict ourselves to a particular class of conformal transformation
for which the conformal factor is given by a power of the module of
the complex scalar field. For a particular value of the exponent, we
prove that the conformally invariant Klein-Gordon action rescales to
the Einstein-Hilbert action with a cosmological constant and with a
source given by a phantom field. The terminology of phantom field
refers to the fact that its kinetic energy term enters in the action
with the opposite sign. As shown below, this is a direct consequence
of the fact that the phantom field is described by the phase of the
complex scalar field. In the recent literature, phantom field has
been proposed to model the observational evidence for accelerating
universe, \cite{Caldwell:1999ew} and \cite{Carroll:2003st}. In this
case, the resulting equation of state parameter is given by
$\omega=p/\rho<-1$ where $p$ is the pressure and $\rho$ the energy
density. In contrast, the usual scalar field theories used to model
the dark energy as the quintessence or the tachyonic scalar fields
lead to an equation of state $\omega\geq -1$ which results as a
consequence of imposing the dominant energy condition. However,
classical tests of cosmology do not seem to exclude the regime
$\omega< -1$, \cite{Caldwell:1999ew}.

The correspondence between conformally invariant complex scalar
field and the phantom field coupled to gravity is also established
at the level of the equations of motion. More precisely, solutions
of the non-linear Klein-Gordon equation together with the vanishing
of the energy-momentum tensor are mapped to solutions of a phantom
scalar field satisfying a massless Klein-Gordon equation and
minimally coupled to gravity with a cosmological constant. The
converse of this mapping is also valid with the advantage that it
offers more possibilities reflected by the freedom of rewriting a
metric as the conformal of another metric. In particular,
considering the metric as the conformal of itself, the same metric
can described the spacetime geometry of both systems.

The clue of this correspondence lies in the conformal invariance of
the generalized Klein-Gordon action. Hence its coupling with
standard Einstein gravity will break the conformal invariance.
However, in three dimensions gravity governed by the Cotton tensor
enjoys the conformal invariance and in this case, the correspondence
is established with the topologically massive gravity
\cite{Deser:1981wh} with a cosmological constant and with a phantom
source.

The plan of the paper is organized as follows. We review the
conformally invariant Klein-Gordon action for a complex scalar field
with a self-interacting potential. We explicitly show that this
action rescales to the action of a phantom scalar field minimally
coupled to gravity with a cosmological constant. This correspondence
is also established at the level of the equations of motion. A
particular attention is devoted to the three-dimensional case for
which the conformally invariant Klein-Gordon action is interpreted
as the source of conformal gravity governed by the Cotton tensor. In
this case, the system is put in equivalence with topologically
massive gravity with a cosmological constant and with a phantom
source. We present some examples of this correspondence. Finally, an
appendix is given in order to explicitly prove the correspondence at
the level of the actions as well their corresponding equations of
motion.

\section{Phantom from conformal invariance}
We consider the conformally invariant Klein-Gordon action for a
complex scalar field in $D$ dimensions,
\begin{eqnarray}
I=\int d^D
x\,\sqrt{-g}\,\Big(\frac{1}{2}(\partial_{\alpha}\Psi)\,(\partial^{\alpha}\Psi^{\star})
+\frac{\xi_{D}}{2}R\,\Psi\Psi^{\star}\nonumber\\
+\lambda \left(\Psi\Psi^{\star}\right)^{\frac{D}{D-2}}\Big),
\label{GKGaction}
\end{eqnarray}
where $R$ represents the scalar curvature of the metric, $\lambda$
is the potential strength and $\xi_D$ is the conformal nonminimal
coupling parameter given by
\begin{eqnarray}
\xi_D=\frac{(D-2)}{4(D-1)}. \label{xiconf}
\end{eqnarray}
It is well-known that for this specific value (\ref{xiconf}), the
action (\ref{GKGaction}) possesses a conformal invariance for which
the implementation on the dynamical fields is given by
\begin{eqnarray}
g_{\mu\nu}\to \Omega^2g_{\mu\nu},\qquad
\Psi\to\Omega^{\frac{2-D}{2}}\Psi. \label{imple}
\end{eqnarray}
The equation of motion obtained by varying this action with respect
to the scalar field yields to a non-linear generalized Klein-Gordon
equation
\begin{eqnarray}
\Box\Psi=\xi_D R\Psi+\frac{2\lambda
D}{(D-2)}\left(\Psi\Psi^{\star}\right)^{\frac{2}{D-2}}\Psi.
\label{nlGKG}
\end{eqnarray}
The variation of (\ref{GKGaction}) with respect to the metric
implies the vanishing of the energy-momentum tensor, i. e.
\begin{eqnarray}
&&T_{\mu\nu}:=\nabla_{(\mu}\Psi\nabla_{\nu )}\Psi^{\star}
+\xi_D\left(g_{\mu\nu}\Box-
\nabla_{\mu}\nabla_{\nu}+G_{\mu\nu}\right)\vert\Psi\vert^2\nonumber\\
&&-g_{\mu\nu}
\left(\frac{1}{2}\nabla_{\sigma}\Psi\nabla^{\sigma}\Psi^{\star}+\lambda
\left(\Psi\Psi^{\star}\right)^{\frac{D}{D-2}}\right)=0.
\label{tracemape}
\end{eqnarray}
For later convenience, we decompose the complex scalar field into
its module and phase as $\Psi=\sqrt{\rho}\,e^{i\theta}$.

We now show that, under a particular conformal transformation, the
action (\ref{GKGaction}) rescales to the action of a phantom scalar
field coupled to gravity with a cosmological constant. Indeed, it is
straightforward to see that in the conformal frame given by
\begin{eqnarray}
g_{\mu\nu}^{\prime}=\rho^{\frac{2}{D-2}}\,g_{\mu\nu}, \label{yamabe}
\end{eqnarray}
the scalar field action (\ref{GKGaction}) becomes, up to a boundary
term,
\begin{eqnarray}
I^{\prime}=\frac{\xi_D}{2}\int d^D
x\,\sqrt{-g^{\prime}}\left[R^{\prime}-2(-\frac{\lambda}{\xi_D})+\frac{1}{\xi_D}
\partial_{\mu}\theta\partial^{\mu}\theta\right].
\label{phantom}
\end{eqnarray}
We recognize the action of a phantom scalar field minimally coupled
to gravity with a cosmological constant. We refer to the dynamical
field $\theta$ as a phantom scalar field since its kinetic term
enters in the action (\ref{phantom}) with the opposite sign (we are
using the mostly plus signature). In the expression (\ref{phantom}),
the cosmological constant $\Lambda$ is expressed in terms of the
potential strength and the conformal nonminimal coupling parameter
(\ref{xiconf}) as
\begin{eqnarray}
\Lambda=-\frac{\lambda}{\xi_D}=-\frac{4\lambda(D-1)}{(D-2)}.
\label{concos}
\end{eqnarray}
The field equations associated to the rescaled action
(\ref{phantom}) read
\begin{subequations}
\label{eqsmotion2}
\begin{eqnarray}
\label{einsteineqs}
G_{\mu\nu}^{\prime}-\frac{\lambda}{\xi_D}g_{\mu\nu}^{\prime}=\frac{1}{\xi_D}\vartheta_{\mu\nu},\\
\nonumber\\
 \label{waveeq}
\Box^{\prime}\theta=0,
\end{eqnarray}
\end{subequations}
where the energy-momentum tensor $\vartheta_{\mu\nu}$ is given by
\begin{eqnarray}
\vartheta_{\mu\nu}=-\partial_{\mu}\theta\partial_{\nu}\theta+\frac{1}{2}g_{\mu\nu}^{\prime}
\,g^{{\prime}{\alpha\beta}}\partial_{\alpha}\theta\partial_{\beta}\theta.
\label{vartheta}
\end{eqnarray}
The correspondence is also valid at the level of the equations of
motion. Indeed, the energy-momentum tensor $T_{\mu\nu}$ defined in
Eq. (\ref{tracemape}) becomes in the new frame (\ref{yamabe})
\begin{eqnarray}
T_{\mu\nu}=\xi_D\rho\Big[G_{\mu\nu}^{\prime}-\frac{\lambda}{\xi_D}g_{\mu\nu}^{\prime}-\frac{1}{\xi_D}
\Big(-\partial_{\mu}\theta\partial_{\nu}\theta \nonumber\\
+\frac{1}{2}g_{\mu\nu}^{\prime}{g^{\prime}}^{\alpha\beta}\partial_{\alpha}\theta\partial_{\beta}\theta\Big)\Big].
\label{rescaltmunu}
\end{eqnarray}
Hence, the vanishing of the energy-momentum tensor (\ref{tracemape})
is equivalent to the Einstein equations with phantom source
(\ref{einsteineqs}). It can also be shown that the imaginary part of
the non-linear Klein-Gordon equation (\ref{nlGKG}) becomes the
massless Klein-Gordon equation for the phantom field (\ref{waveeq})
while its real part implies
\begin{eqnarray}
R^{\prime}=\frac{2\lambda
D}{(2-D)\xi_D}-\frac{1}{\xi_D}\partial_{\sigma}\theta\partial^{\sigma}\theta,
\label{scalarcurvaturee}
\end{eqnarray}
which is a consequence of the Einstein equations
(\ref{einsteineqs}).

In sum, we have shown that starting from a complex scalar field
$\Psi=\sqrt{\rho}\,e^{i\theta}$ and a metric $g_{\mu\nu}$ satisfying
the equations (\ref{nlGKG}-\ref{tracemape}), the rescaled metric
(\ref{yamabe}) together with the phase $\theta$ are solutions of the
Einstein equations with phantom source (\ref{eqsmotion2}). The
converse is also valid with the advantage that it offers more
possibilities of mapping. Indeed, there exists a freedom of
rewriting the metric solution $g_{\mu\nu}^{\prime}$ as the conformal
of another metric and each of these choices will yield to a
different solution of the Klein-Gordon action by engineering
inverse. In particular, the same metric can described the spacetime
geometry of both system provided the complex scalar field is only
giving by a phase term, $\Psi=e^{i\theta}$.

\subsection{Three-dimensional conformal gravity}

The clue of this correspondence is due to the conformal invariance
of the Klein-Gordon action (\ref{GKGaction}) and, hence the coupling
of this action to the Einstein gravity will explicitly break this
invariance. In three dimensions, the action (\ref{GKGaction}) can be
seen as the source of conformal gravity governed by the Cotton
tensor and consequently the full system enjoys the conformal
invariance. In this case, the three-dimensional field equations
become
\begin{subequations}
\label{lk}
\begin{eqnarray}
\label{waveeqrealcg} \Box\Psi=\frac{1}{8}R\Psi+6\vert\Psi\vert^2\Psi,\\
\label{Tmunueqscg} C_{\mu\nu}=\kappa T_{\mu\nu},
\end{eqnarray}
\end{subequations}
where the energy-momentum tensor $T_{\mu\nu}$ is given by the
expression (\ref{tracemape}) with $D=3$ and $C_{\mu\nu}$ is the
Cotton tensor defined by \footnote{Our conventions are the
following: the signature is $(-++)$, $\epsilon_{012}=+1$, the
Riemann tensor is $R_{\mu\nu\rho}^{\,\;\quad\sigma}=+
\partial_{\nu}\Gamma^{\sigma}_{~\mu\rho} - \ldots$ and the Ricci tensor is
$R_{\mu\nu}=+\partial_\alpha\Gamma^\alpha_{~\mu\nu} - \ldots$.}
\begin{eqnarray}
C^{\mu\nu}=\frac{1}{\sqrt{-g}}\epsilon^{\mu\alpha\beta}
D_{\alpha}\left(R_{\beta}^{~\nu}-\frac{1}{4}\delta_{\beta}^{~\nu}R\right).
\label{eq:cotton}
\end{eqnarray}
This tensor is symmetric, identically conserved and traceless.
Hence, since the Cotton tensor enjoys the conformal symmetry, it is
easy to see that in the frame
$$
g^{\prime}_{\mu\nu}=\rho^2\,g_{\mu\nu},
$$
the equations of three-dimensional conformal gravity with source
(\ref{lk}) become a massless Klein-Gordon equation for the phantom
field, i.e. $\Box^{\prime}\theta=0$, together with
\begin{eqnarray}
\frac{1}{\mu}C_{\mu\nu}^{\prime}+G_{\mu\nu}^{\prime}- 8\lambda
g_{\mu\nu}^{\prime}=8
\left[-\partial_{\mu}\theta\partial_{\nu}\theta+\frac{1}{2}g_{\mu\nu}^{\prime}
\partial_{\sigma}\theta\partial^{\sigma}\theta\right],
\label{tmg}
\end{eqnarray}
where $\mu=-\kappa/8$. We recognize the equations describing a
phantom scalar field acting as a source for topologically massive
gravity with cosmological constant. To be more precise, in order to
make contact with topologically massive gravity the signs of the
topological mass $\mu$ and the dimensionless parameter $\kappa$ must
be different. This ambiguity is irrelevant when the constants of the
problem are not tied since we are just mapping solution of a system
to solution of another one.

\subsection{Examples}
In what follows, we provide three explicit examples of the
correspondence. The first two ones are given by the Siklos spacetime
which provides a natural setup of the machinery described before
since this spacetime is conformally related to a pp wave geometry.
The third example is an ultra static wormhole solution of the
Einstein equations with a phantom source.

\subsubsection{The pp wave metric and the Siklos spacetime}
We propose to solve the Einstein equations with a phantom source
(\ref{eqsmotion2}) in $D$ dimensions with a geometry given by
\begin{eqnarray}
ds^2=\frac{(D-2)^2}{8\lambda
y^2}\Big[-F(u,y)du^2-2dudv+dy^2\nonumber\\+dx_1^2+\cdots+dx_{D-3}^2
\Big]. \label{adswaveansatz}
\end{eqnarray}
This metric belongs to the class of Siklos spacetimes which are the
only non-trivial Einstein spaces conformal to non-flat pp waves.
These metrics are of importance in the study of propagation of
gravitational waves in presence of a cosmological constant.
Recently, real scalar fields nonminimally coupled to this geometry
have been derived in three dimensions \cite{ABH}.

We assume that the null Killing field
$k^{\mu}\partial_{\mu}=\partial_v$ is also a symmetry of the phantom
scalar field, that means $\theta=\theta(u,y,\vec{x})$. In this case,
it is easy to see that the only non-vanishing component of the
Einstein tensor with cosmological constant is the one along the
retarded time $(uu)$ that reads
$$
G_{uu}-\frac{\lambda}{\xi_D}
g_{uu}=\frac{1}{2y}\Big((\partial_{yy}F) y-(D-2)\partial_{y}F\Big).
$$
Moreover, the remaining Einstein equations reduce to the vanishing
of the energy-momentum tensor $\vartheta_{\mu\nu}$ which in turn
imply that the function $\theta$ depends only on the retarded time,
$\theta=\theta(u)$. Finally, the integration of the $uu-$component
of the Einstein equations yields for $D\not=3$
\begin{eqnarray}
F(u,y)=\frac{1}{\xi_D(D-3)}(\partial_u\theta)^2y^2+C(u)y^{D-1},
\label{sold}
\end{eqnarray}
while in three dimensions the structural function is given by
\begin{eqnarray}
F(u,y)=-8(\partial_u\theta)^2y^2\ln y+4
(\partial_u\theta)^2y^2+C(u)y^2.\label{sol3d}
\end{eqnarray}
In both cases, the real scalar field $\theta$ is an arbitrary
function of the retarded time and $C$ is an undetermined integration
function of the retarded time.

By engineering inverse, we conclude that the following complex
scalar field
\begin{eqnarray}
\Psi(u,y)=\left[\frac{(D-2)}{8\lambda y^2}\right]^{\frac{D-2}{4}}
\,e^{i\theta(u)}, \label{wcf}
\end{eqnarray}
together with the pp wave metric
\begin{eqnarray}
ds^2=-F(u,y)du^2-2dudv+dy^2+\sum_{i=1}^{D-3}dx_i^2, \label{ppwave}
\end{eqnarray}
satisfy the non-linear Klein-Gordon equation with vanishing
energy-momentum tensor (\ref{nlGKG}-\ref{tracemape}).

\subsubsection{Conformal gravity in three dimensions with Siklos
spacetime}
We now consider the equations of topologically massive gravity with
cosmological constant with a source given by a phantom scalar field
(\ref{tmg}). As previously, we look for a geometry whose line
element is given by the Siklos spacetime (\ref{adswaveansatz}),
\begin{eqnarray}
ds^2=\frac{1}{8\lambda y^2}\Big[-F(u,y)du^2-2dudv+dy^2 \Big].
\label{adswaveansatz3}
\end{eqnarray}
In this case, the only non-vanishing component of the left hand side
of the equations (\ref{tmg}) is the one along the retarded time
$uu$. As previously, the vanishing of the remaining component of the
energy-momentum tensor implies that the scalar field $\theta$
depends only on the retarded time $u$, and hence the component $uu$
of the equations (\ref{tmg}) becomes
\begin{eqnarray*}
-\frac{8\sqrt{2\lambda}}{\kappa}y\,\partial_{yyy}F+\frac{1}{2y^2}
\left(y^2\,\partial_{yy}F-y\,\partial_{y}F\right)=-8\left(\partial_u\theta\right)^2.
\label{er}
\end{eqnarray*}
For a coupling constant $2\lambda\not=(\kappa/16)^2$ , the
integration of this equation yields
\begin{eqnarray}
\label{FF}
F(u,y)&&=C_1(u)\,y^2+C_2(u)\,y^{\frac{\kappa\sqrt{2}}{32\sqrt{\lambda}}+1}\\
&&-\frac{8\kappa\,y^2\ln
y(\partial_u\theta)^2}{(\kappa-16\sqrt{2\lambda})}+ \frac{4\kappa
y^2(\partial_u\theta)^2}{(\kappa-16\sqrt{2\lambda})^2}(\kappa-48\sqrt{2\lambda})\nonumber
\end{eqnarray}
where $C_1$ and $C_2$ are two undetermined functions of the retarded
time. For the special coupling $2\lambda=(\kappa/16)^2$, the
solution is
\begin{eqnarray}
F(u,y)=C_1(u)\,y^2+C_2(u)\,y^2\ln y+4(\partial_u\theta)^2y^2(\ln
y)^2.\nonumber\\
\label{fff}
\end{eqnarray}
Hence, using the correspondence one concludes that for a coupling
constant $2\lambda\not=(\kappa/16)^2$, the complex scalar field
given by
\begin{eqnarray}
\Psi=\frac{1}{\sqrt{\sqrt{8\lambda}\,y}}\,e^{i\theta(u)},
\end{eqnarray}
together with the pp wave geometry for which the metric function
$F(u,y)$ is given by (\ref{FF}) are solutions to the
three-dimensional conformal gravity equations (\ref{lk}). For the
special coupling $2\lambda=(\kappa/16)^2$, the metric function of
the pp wave geometry is given by (\ref{fff}) and the complex scalar
field is
$$
\Psi=\sqrt{\frac{8}{\kappa y}}\,e^{i\theta(u)},
$$
where $\theta$ is an arbitrary function of the retarded time.

\subsubsection{Massless wormholes in four dimensions}
In four dimensions, there exists a regular ultra static wormhole
solution of the Einstein equations  without cosmological constant,
i.e. equations (\ref{eqsmotion2}) with $\lambda=0$. The metric
solution in isotropic coordinates reads
\cite{Gibbons:1996pd}-\cite{Gibbons:2003yj},
\begin{eqnarray}
ds^2=-dt^2+\frac{c^2}{4}\left(1+\frac{1}{r^2}\right)^2(dx^2+dy^2+dz^2),
\label{wm}
\end{eqnarray}
with $r^2=x^2+y^2+z^2$ and, the phantom scalar field $\theta$ is
given by
\begin{eqnarray}
\theta=\frac{\sqrt{3}}{3}\arctan\left(\frac{r^2-1}{2r}\right).
\label{theta}
\end{eqnarray}
The spatial sections have the form of an Einstein-Rosen bridge
joining two isometric regions each with vanishing ADM mass. The
wormhole metric (\ref{wm}) can be seen as the conformal of itself
and, hence the complex scalar field defined by
\begin{eqnarray}
\psi=e^{i\frac{\sqrt{3}}{3}\arctan\left(\frac{r^2-1}{2r}\right)},
\label{solcc}
\end{eqnarray}
with the wormhole metric (\ref{wm}) are also solutions of the
original equations (\ref{nlGKG}-\ref{tracemape}) without
self-interacting potential $\lambda=0$.

\section{Discussion}
Here, we have shown that a phantom field coupled to gravity can be
put in equivalence with a conformally invariant complex scalar
field. The spacetime geometry of both systems are conformally
related with a factor proportional to a power of the module of the
complex scalar field while its phase plays the role of the phantom
field. This last fact legitimates the reason for which the phantom
kinetic term enters in the action with the opposite sign. Indeed,
being purely imaginary its kinetic expression arises with a negative
sign. In the current literature, the phantom field and its peculiar
dynamics have been introduced to model the observed acceleration of
the scale factor of the universe. In view of the correspondence
reported here, an interesting and natural work will consist to
analyze the cosmological consequences of interpreting the phantom
field as a conformally invariant complex scalar field.

We have also established that the correspondence allows to map
solutions of the conformally action to those of the phantom field
minimally coupled to gravity. The converse is also valid with the
advantage that it offers more possibilities of mapping. For a metric
and a phantom field satisfying the Einstein equations with a
massless Klein-Gordon equation, one can generate a wide variety of
solutions of the non-linear Klein-Gordon equation with vanishing
energy-momentum tensor. All these possibilities are due to the
freedom of rewriting a metric as the conformal of another metric,
and each of these choices will yield to a different solution of the
non-linear Klein-Gordon equation with $T_{\mu\nu}=0$. However, in
all these cases the complex scalar field solutions will have the
same phase and, in some sense they will belong to a same class of
equivalence. In particular, it is intriguing that the same metric
can described both systems provided that the complex scalar field is
only given by a phase term.

Finally, to conclude it would be desirable to have a physical
interpretation of this correspondence.

{\bf Acknowledgments.-} We thank E. Ay\'on-Beato, J. Gomis, C.
Mart\'{\i}nez,  R. Troncoso and J. Zanelli for useful discussions.
This work is partially supported by grants 1051084 from FONDECYT.
Institutional support to the Centro de Estudios Cient\'{\i}ficos
(CECS) from Empresas CMPC is gratefully acknowledged. CECS is a
Millennium Science Institute and is funded in part by grants from
Fundaci\'{o}n Andes and the Tinker Foundation.

\appendix

\section{\label{app:conft} Conformal transformations}
We consider a conformal transformation of the metric $g_{\mu\nu}$ as
\begin{eqnarray*}
{g}_{\mu\nu}^{\prime}=\rho^{\frac{2}{D-2}}\,g_{\mu\nu},\qquad
\mbox{or}\qquad
{g}_{\mu\nu}=\rho^{\frac{2}{2-D}}\,g_{\mu\nu}^{\prime}\label{conftrans}
\end{eqnarray*}
Under this transformation, the operator
$\Box=\nabla_{\mu}\nabla^{\mu}$ becomes
\begin{subequations}
\label{box}
\begin{eqnarray*}
{\Box}^{\prime}=\frac{1}{\rho^{\frac{2}{D-2}}}\Box+
\frac{1}{\rho^{\frac{D}{D-2}}}g^{\alpha\beta}(\partial_{\alpha}\rho)\partial_{\beta},\\
\label{boxinv}
\Box=\rho^{\frac{2}{D-2}}\left[\Box^{\prime}-\frac{1}{\rho}
{g^{\prime}}^{\alpha\beta}(\partial_{\alpha}\rho)\partial_{\beta}\right].
\end{eqnarray*}
\end{subequations}
The scalar curvature becomes
\begin{subequations}
\label{scaR}
\begin{eqnarray*}
{R}^{\prime}=\frac{1}{\rho^{\frac{2}{D-2}}}\left[R+\frac{(D-1)}{(D-2)}\left(-2\frac{\Box\rho}{\rho}
+\frac{g^{\alpha\beta}\nabla_{\alpha}\rho\nabla_{\beta}\rho}{\rho^2}\right)\right]\\
\label{scaR2}
R=\rho^{\frac{2}{D-2}}\left[R^{\prime}+\frac{(D-1)}{(D-2)}
\left(\frac{2\,\Box^{\prime}\rho}{\rho}
-3\frac{{g^{\prime}}^{\alpha\beta}\nabla_{\alpha}\rho\nabla_{\beta}\rho}{\rho^2}\right)\right]
\end{eqnarray*}
\end{subequations}
and the Einstein tensor yields
\begin{eqnarray*}
{G}_{\mu\nu}^{\prime}=&&G_{\mu\nu}-\frac{\nabla_{\mu}\nabla_{\nu}\rho}{\rho}
+\frac{(D-1)}{(D-2)}\frac{\nabla_{\mu}\rho\nabla_{\nu}\rho}{\rho^2}\nonumber\\
&&+g_{\mu\nu}\left[\frac{\Box\rho}{\rho}-\frac{(D-1)}{2(D-2)}\frac{\nabla_{\sigma}
\rho\nabla^{\sigma}\rho}{\rho^2}\right], \label{einsteintensor}
\end{eqnarray*}
or equivalently
\begin{eqnarray*}
{G}_{\mu\nu}=&&{G}_{\mu\nu}^{\prime}+\frac{\nabla^{\prime}_{\mu}\nabla^{\prime}_{\nu}\rho}{\rho}
-\frac{D-3}{D-2}\frac{\nabla_{\mu}\rho\nabla_{\nu}\rho}{\rho^2}\\
             &&+g_{\mu\nu}^{\prime}\left[-\frac{\Box^{\prime}\rho}{\rho}+
             \frac{(3D-7)}{2(D-2)}\frac{{g^{\prime}}^{\alpha\beta}\nabla_{\alpha}
\rho\nabla_{\beta}\rho}{\rho^2}\right].
\end{eqnarray*}


\end{document}